\documentclass[rnote]{aa}

\usepackage{graphicx}
\usepackage{txfonts}
\usepackage{natbib}
\begin{document}

	\title{The properties of the putative pulsar associated with IGR J18135$-$1751/HESS J1813$-$178}

   \author{A. J. Dean
          \inst{1}
          \and
          A. B. Hill\inst{1}
          }

   \offprints{A. J. Dean}

   \institute{School of Physics \& Astronomy, University of Southampton, UK\\
              \email{ajd@astro.soton.ac.uk}
              }

   \date{Received; accepted}

 
  \abstract
   {We investigate the possible theoretical properties of the putative pulsar associated with the pulsar wind nebula IGR J18135$-$1751/HESS J1813$-$178 based upon recent $\gamma$-ray observations and archival multi-wavelength observations.}
   {We show that when using the standard equations for magnetic dipole radiation with recent soft $\gamma$-ray observations leads to deriving an extreme set of parameters (magnetic field, period and spin down rate) for the putative pulsar.  Alternative scenarios that generate more typical parameter values are explored.}
   {The properties of the putative pulsar are calculated assuming that the 20--100 keV luminosity corresponds to 1\% of $\dot{E}$, that the source is 4.5 kpc away, and that the pulsar age is 300 yrs.  This gives $P =$ 0.55 s, $\dot{P}$ = 3$\times$10$^{-11}$ss$^{-1}$, and $B$ = 1.28$\times$10$^{14}$ G.  This is a very extreme set compared to the population of known pulsars in PWN systems.  Using the equations for magnetic dipole losses makes it possible to adjust the initial assumptions to see what is required for a more reasonable set of pulsar parameters.}
   {The current measured properties for IGR J18135$-$1751/HESS J1813$-$178 (i.e. luminosity, distance, and age) result in extreme properties of the unseen pulsar within the PWN.  The simplest method for achieving more reasonable properties for the pulsar is to decouple the spin-down age of the pulsar from the actual age for the system.}
   {}


   \maketitle
%

\section{Introduction}

HESS J1813$-$178 was discovered in the HESS galactic plane survey %
\citep{2006ApJ...636..777A} and was seen to be one of the more compact TeV
sources, with a radius of $\sim$2$\arcmin$. The source has a hard power law
spectrum with $\Gamma$ $\sim$2.09, and a corresponding flux of $\sim$14.2~$%
\times$~10$^{-12}$ photons cm$^{-2}$ s$^{-1}$ above 200 GeV. Originally the
source was classified as unidentified and located at RA = 18$^{h}$ 13$^{m}$
37$^{s}$.9 and Dec = $-$17$^{\circ}$ 50$\arcmin$ 34$\arcsec$ with a positional
accuracy in the range 1--2$\arcmin$. However the study of previously
unpublished archival radio data showed the TeV source to be coincident
with the faint radio emission from part of the shell structure of a
supernova remnant (SNR G12.8$-$0.0), which lies 8$\arcmin$ above the Galactic
plane and in the vicinity of the bright star forming region W33 %
\citep{2005ApJ...629L.105B}.  However, no radio pulsar was (or has yet been) found in the
vicinity \citep{2007ApJ...665.1297H}.

Likewise a study of previously unpublished ASCA X-ray data revealed a bright
non-thermal 2--10 keV X-ray source, AX J1813$-$178 \citep{2005ApJ...629L.105B}%
. The ASCA angular resolution does not permit the X-rays to be associated
with either the SNR shell or a putative compact object near the centre. The
X-ray spectrum was found to be quite hard, with the emission extending to 10
keV and a strong cutoff below 2 keV. Typical parameters associated with the
X-ray emission are N$_{H}$ $\sim $10$^{23}$cm$^{-2}$, and $\Gamma $ $\sim $%
1.83. The 2--10 keV unabsorbed flux is 7~$\times $~10$^{-12}$ erg cm$^{-2}$ s%
$^{-1}$, corresponding to a 2--10 keV luminosity L$_{X}$ $\sim $1.7~$\times $%
~10$^{34}$ erg s$^{-1}$, for a distance of 4.5 kpc to the source as
suggested from the measured N$_{H}$ value. The hard power law spectral index
is compatible with a pulsar/PWN system, although a careful search in the
ASCA high bit-rate GIS data by \citet{2005ApJ...629L.105B} did not
detect a pulsed signal between 4 and 8 keV for periods in the range 125ms to
1000s. The small angular size of SNR G12.8$-$0.0 ($\phi $ $\sim $2.5$\arcmin$)
suggests youth. At 4.5 kpc the radius will be $\sim $1.6 pc and, if the SNR
is still freely expanding with a typical velocity of 5000 km s$^{-1}$, it
will only be a few hundred years old. We may anticipate that the putative
pulsar has a period of less than 124ms, typical of many of the pulsar/PWN X-
and $\gamma$-ray emitting systems discovered to date \citep{2002A&A...387..993P}.

XMM-Newton observations \citep{2007A&A...470..249F} have revealed a highly
absorbed (N$_{H}$ $\sim $10$^{23}$ cm$^{-2}$) non-thermal point-like object
coincident with the ASCA source inside the radio shell and having a faint
tail towards the north-east resembling a PWN system. The basic scenario is
essentially confirmed by recent \textit{Chandra }observations of HESS J1813$-
$178 \citep{2007ApJ...665.1297H}. The \textit{Chandra} image resolves the
ASCA source into diffuse X-ray emission and a point source. The diffuse
emission generally fills the radio shell and peaks towards the point source
emission, which lies within, but slightly offset from the centroid of the
SNR by about 20$\arcsec$. Spectra from each morphological region are well
characterized by an absorbed power law model associated with non-thermal
emission. The best fit photon index for the nebular flux is $\Gamma $ $\sim $%
1.3 with N$_{H}$ $\sim $9.8$\times$ 10$^{22}$ cm$^{-2}$, and the point source having
a similar spectrum, also at $\Gamma $ $\sim $1.3. These values are typical
of other energetic young pulsars. For a distance of 4.5 kpc the \textit{%
Chandra} results correspond to a 2--10 keV luminosities of the putative
pulsar and PWN of L$_{PSR}$ $\sim $3.2~$\times $~10$^{33}$ erg s$^{-1}$ and L%
$_{PWN}$ $\sim $1.4~$\times $~10$^{34}$ erg s$^{-1}$.

\subsection{Gamma-ray observations}
\citet{2005ApJ...629L.109U} report the discovery of a soft $\gamma$-ray source
IGR J18135$-$1751, detected by the IBIS telescope on INTEGRAL, coincident
with the ASCA and HESS emissions within the errors of the instrument. The
source is persistent and has a 20--100 keV luminosity of L$_{S\gamma }$ $%
\sim $7.2~$\times$~10$^{34}$ erg s$^{-1}$ if situated at 4.5 kpc. Due to the
lack of X- and $\gamma $-ray variability, the radio morphology and ASCA
spectrum the authors interpret this source as a PWN system embedded in its
supernova remnant. Using the data set of the recently released 3$^{rd}$
IBIS/ISGRI survey catalogue \citep{2007ApJS..170..175B}, we measure the
source flux to be 2.75$_{-0.60}^{+0.83}$~$\times$~10$^{-12}$ erg cm$^{-2} $ s%
$^{-1}$ corresponding to a source luminosity (at 4.5 kpc) of L$_{S\gamma } $ 
$\simeq $ 6.7~$\times$~10$^{34}$ erg s$^{-1}$. The spectral index of the
power law interpretation of the spectrum is well described by $\Gamma $ =
1.8. If we look at the luminosities as a function of $\dot{E}$ all PWN seen in the IBIS/ISGRI survey catalogue of \citet{2007ApJS..170..175B} we find a distribution centred on 1\% as seen in Figure~\ref{fig:histo}. Taking the 20--100 keV luminosity to be roughly 1\% of the spin down power, then we may expect to find a pulsar in IGR J18135$-$1751 with $\dot{E}$ $\sim$ 6.7~$\times$%
~10$^{36}$ erg s$^{-1}$.

In discussing the TeV emission the HESS team suggest a model in which the
X-rays are produced by synchrotron radiation by electrons in a PWN and the
TeV photons arise from Inverse Compton (IC) scattering dust IR photons, the
cosmic microwave background and ambient starlight. IC scattering from the
Cosmic Microwave Background alone is not sufficient, so that in this context
the proximity of HESS J1813$-$178 to W33 would be a definite advantage.
Taking the XMM, INTEGRAL and HESS spectral data, \citet{2007A&A...470..249F}
find that the scenario, in which the VHE and X-ray emitting electrons belong
to the same population that originate in a single central object, as
suggested by \citet{2005ApJ...629L.109U} provides a good fit to the spectral
energy distribution.

\section{The Properties of the Putative Pulsar}

The morphology and spectral emission from the central region of G12.8$-$0.0
indicate the presence of a pulsar, which is powering the system. \textit{%
Chandra} currently provides a compelling case that the high-energy emission
from G12.8$-$0.0 is derived from the spin down of a young rotation-powered
pulsar, all that remains is the discovery and characterization of the
pulsar. If we assume that this pulsar is currently slowing down through the
loss of rotational energy at a rate $\dot{E}$ through magnetic dipole
radiation, for which the braking index $n=3$, we can derive expressions
relating $\dot{E}$, the period of the pulsar $P$, the period derivative $\dot{P%
}$ and the characteristic age $\tau$ as follows \citep{Padmanabhan...text, Bower...text}:

\begin{eqnarray}
\dot{E} & = & \frac{4\pi ^{2}I\dot{P}}{P^{3}} \ \ \ \mathrm{erg \ s}^{-1}
\label{equ: edot} \\
B & = & 3.2\times10^{19} (P \dot{P})^{0.5} \ \ \ \mathrm{Gauss}
\label{equ: b} \\
\tau & = & \frac{P}{2\dot{P}} \ \ \ \mathrm{s}  \label{equ: tau}
\end{eqnarray}

\noindent where $I$ $\simeq $ 10$^{45}$ g cm$^{2}$ is the moment of inertia of the
neutron star. Eliminating $\dot{P}$ between Equations~\ref{equ: edot}~\&~\ref%
{equ: tau} we derive:

\begin{eqnarray}
\tau P^{2}\dot{E}=1.9\times 10^{46} \ \ \ \mathrm{erg \ s^{2}}.
\end{eqnarray}

\noindent Thus for a given value of $\dot{E}$ the product of $\tau $~$\times $~P$^{2}$
is fixed.  The 3$^{rd}$ IBIS/ISGRI survey catalogue of \citet{2007ApJS..170..175B} includes nine pulsar wind nebula systems with measured pulsar properties, although not all are labelled as being such; these include:
\begin{itemize}
	\item Crab
	\item	PSR B0540$-$69.3 (not listed as PWN in \citet{2007ApJS..170..175B})
	\item	Vela Pulsar
	\item	PSR B1509$-$58	(not listed as PWN in \citet{2007ApJS..170..175B})
	\item	PSR J1617$-$5055	(not listed as PWN in \citet{2007ApJS..170..175B})
	\item	IGR J17475$-$2822	(not listed as PWN in \citet{2007ApJS..170..175B})
	\item	PSR J1811$-$1926
	\item SNR 021.5$-$00.9
	\item	AX J1846.4$-$0258
\end{itemize}
The luminosity distribution of these systems, as measured by INTEGRAL, in the 20--100 keV band as a percentage of $\dot{E}$ is shown in the histogram shown in Figure~\ref{fig:histo}.  From the histogram it can be seen that an average conversion efficiency of spin down energy into soft $\gamma$-rays of $\epsilon $ $\approx$ 1\% is reasonable.  Hence for IGR J18135$-$1751 with a luminosity of \ L$_{S\gamma }$ $\simeq $ 6.7~$\times $~10$^{34}$ erg s$^{-1}$ as measured by INTEGRAL leads to a value of $\dot{E}$ = 6.7~$\times $~10$^{36}$erg s$^{-1}$.  Taking the distance, $d$, to this source to be 4.5kpc, the age of the pulsar to be that of a SNR freely expanding at $\sim$5000 kms$^{-1}$ ($\sim $300 yr) it is possible to derive the following numerical values for the period, the period derivative and the surface magnetic field B of the pulsar:

\begin{figure}[b]
\centering
\includegraphics[width=0.95\linewidth, clip]{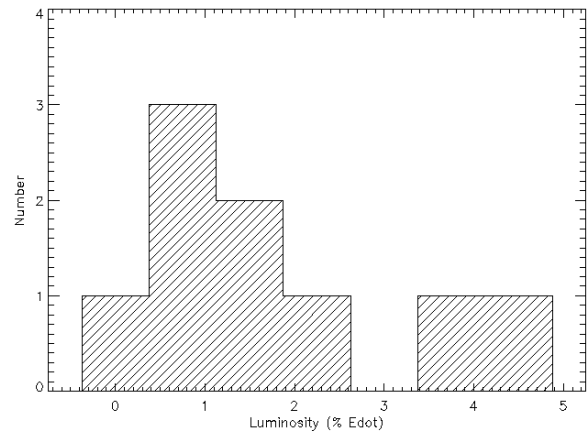}
\caption{Histogram of the soft $\gamma$-ray (20--100 keV) luminosity as a percentage $\dot{E}$ for the 9 PWN systems detected in the 3rd IBIS/ISGRI survey catalogue of \citet{2007ApJS..170..175B} for which the pulsar characteristics are known.}
\label{fig:histo}
\end{figure}

\begin{eqnarray}
P & = & 0.55\frac{\varepsilon _{1}^{0.5}I_{45}^{0.5}}{d_{4.5}\tau
_{300}^{0.5}} \ \ \ \mathrm{s} \\
\dot{P} & = & 3\times 10^{-11}\frac{\varepsilon _{1}^{0.5}I_{45}^{0.5}}{%
d_{4.5}\tau_{300}^{1.5}} \ \ \ \mathrm{ss^{-1}} \\
B & = & 1.28\times 10^{14}\frac{\varepsilon _{1}^{0.5}I_{45}^{0.5}}{%
d_{4.5}\tau_{300}} \ \ \ \mathrm{Gauss}.
\end{eqnarray}

\noindent Where the subscripts represent the normalization of the parameters. i.e. $%
\tau _{300}$ = $\tau /$(300 yr), $\varepsilon _{1}=\varepsilon /(1\%),$ $I%
_{45} = I/$(10$^{45}$g cm$^{2}$), and $d_{4.5} = d/$(4.5 kpc).

The values of the parameters $P$, $\dot{P}$, and $B$ thus derived for a 300
year old pulsar situated at 4.5 kpc constitute an unusual set, making this
object a very extreme member of the group of high B field rotation pulsars %
\citep{2006csxs.book..279K, 2004IAUS..218..225G, 2002ApJ...579L..25C}. The
period is very long for such a young pulsar, and the spindown rate extremely
high, leading to a unusually high magnetic field attached to the neutron
star. The closest known example of a young pulsar/PWN system with a high
magnetic field and fast spindown is PSR J1846$-$0258 in Kes 75 %
\citep{2007McBride}. The equivalent values for the PSR J1846$-$0258 system
are: $P =$ 0.326 seconds; $\dot{P}$ = 7.1~$\times $~10$^{-12}$ ss$^{-1}$; $B =$
4.9~$\times $~10$^{13}$; $\tau $ = 730 yr and $\dot{E}$ = 8.1~$\times $~10$%
^{36}$ erg s$^{-1}$. The above set of parameters for IGR J18135$-$1751 are
far more extreme, and whilst they may be considered possible, they must be
considered unlikely. The situation is exacerbated by the fact that for PSR
J1846$-$0258 the value of the conversion efficiency to soft $\gamma$-rays may
be extremely high, with $\varepsilon$ possibly as high as $\sim$5\% (see Figure~\ref{fig:histo}). Such a high
value for IGR J18135$-$1751 would mean a lower value for $\dot{E}$ making
the product of $\tau $~ $\times ~P^{2}$ even larger, so that for a fixed
age of 300 years we require a longer pulsar period and correspondingly
larger values for $\dot{P}$ and $B$.

One approach to generate a more reasonable set of pulsar characteristics is
to change the values of the system parameters such as the assumed distance,
the age of the pulsar and the conversion efficiency from $\dot{E}$ to
observed $\gamma $-ray luminosity so as to reduce the pulsar's period and
magnetic field parameters to more typical values.

This aim may be achieved through an increase in the value of $\dot{E}$ by
assuming a larger distance or a lower value of $\varepsilon $. Alternatively
an increase in the age $\tau $\ of the pulsar will produce a similar effect.
Increasing the value of $\dot{E}$ by two orders of magnitude makes the value
of $\dot{E}$ = 6.7~$\times $~10$^{38}$erg s$^{-1}$, which is somewhat more
than the highest value of known soft $\gamma$-ray emitting pulsar/PWN systems
(Crab $\sim $4.6~$\times $~10$^{38}$erg s$^{-1}$, and PSR J0540$-$6919 $\sim $
1.48~$\times $~10$^{38}$erg s$^{-1}$). To do this we require that either $%
\varepsilon $ = 0.01\%, making IGR J18135$-$1751 a Vela-like system %
\citep{2007Ap&SS.309..215H}, or the source be removed to an impossible
distance of 45 kpc. A change of two orders of magnitude in the value of $%
\dot{E}$ creates a pulsar system with less extreme characteristics: a period
of 55ms; a period derivative of 3~$\times $~10$^{-12}$ss$^{-1}$; and a
magnetic field of 1.28~$\times $~10$^{12}$ Gauss. However such a
configuration itself stretches the current known limits of the pulsar spin
down power and couples it with an exceptionally low value of $\varepsilon $,
again necessitating a very unusual system. To produce a comparable effect,
using the age of the pulsar as a free parameter we have to make the age of
the system 30,000 years, definitively decoupling the putative pulsar from
SNR G12.8$-$0.0.

Constraints on the distance of the object have been derived from HI
absorption measurements as well as the strong absorption found in the X-ray
data. (Brogan et al, 2005). G12.8$-$0.0 lies in the projected vicinity of the
W33 region, a complex of HII regions and massive star forming region. The
W33 complex has been well studied and its distance has been estimated to be 
$\sim$4kpc (See Funk et al., 2007 for a discussion). The N$_{H}$ value
($\sim$10$^{23}$ cm$^{-2}$) is somewhat higher than the total column
density in this direction indicating that the X-ray source is embedded in a
dense environment and/or possibly located behind W33 making the distance
estimate to the source $\gtrsim$4 kpc. In the context of an age constraint,
the small angular size of G12.8$-$0.0 strongly suggests a young SNR, and
clearly the estimated age is coupled to the distance value. However, since
the swept up mass would be still a small fraction of the ejected mass if the
distance is less than $\sim$10 kpc, so that free expansion is justified
within this distance range. At greater distances the SNR is likely to have
entered the Sedov-Taylor phase, for which the age estimate will still
dictate a young system, if G12.8$-$0.0 is expanding into a medium of typical
density. See Brogan et al (2005), who on this basis estimate the age range
to be 285--2500 years.

Based on the assumption that the value of $\varepsilon $ is close to the \
1\% observed for other soft -$\gamma -$ray emitting pulsar/PWN systems, then
it is impossible to obtain a reasonable set of self-consistent parameters,
based on a system driven by magnetic dipole energy losses, linking the
pulsar period, the period derivative and the magnetic field of the rotating
neutron star, without stretching the both distance and age values to the
extreme limits based on current estimates. For example taking an extreme age
of 2500 years, then an unlikely distance of $\sim$15 kpc is required
to provide a reasonable set of values. 

All the morphological evidence compiled from observations of SNR
G12.8$-$0.0/IGR J18135$-$1751/HESS J1813$-$178 points to the presence of a
young and energetic pulsar, which is powering a PWN system. The probability
of a chance positional coincidence of the hard spectrum X-ray source within
the radio shell of  the G12.8$-$0.0 composite SNR, at the location of the TeV $%
\gamma $-ray source HESS J1813$-$178 is extremely low.  Using a simple monte-carlo approach this probability can be estimated; if we pessimestically assume that all 45 HESS catalogue sources, 421 INTEGRAL catalogue sources and 265 known SNRs \citep{SNRcat} lie within 5$^{\circ}$ of the galactic plane, that SNRs are typically 15$\arcmin$ in size and that all HESS and INTEGRAL detections have a poor location accuracy of 3$\arcmin$ then the probability of chance coincidence is 10$^{-5}$--10$^{-6}$.  However we have seen
that if a pulsar was indeed born at the time of this young supernova remnant
with a period extremely short compared to its current value then it is
extremely difficult to reconcile the properties of the putative pulsar with
the likely age and likely distance to the object. Only by evoking an extreme
neutron star configuration judged by current observational data or
stretching the bounds of both the age and distance is it possible to create
a reasonable physical rotating neutron star system.

All the above discussion assumes explicitly that magnetic dipole losses are
the source of power for the system and that the associated pulsar was born
with a spin period that was considerably faster than the current and unknown
value. By decoupling the spin down age of the putative pulsar from the
actual age of the system, the dilemma may be resolved. AX J1813$-$178 /IGR
J18135$-$1751 may be a system containing a pulsar that was created with a
spin period not significantly different from the present value. Such a
scenario is not unprecedented. The discovery of a 65 millisecond pulsar (PSR
J1811$-$1925) in the supernova remnant G11.2$-$0.3 with ASCA \citep{1997ApJ...489L.145T} sets a precedent. The object was known to be a composite supernova remnant
having an extended shell component and a compact plerionic component. The
age of the SNR was estimated on the basis of its likely association with the
supernova A.D. 386. The discovery of the pulsar and subsequent detailed
Chandra X-ray observations (Kaspi et al., 2001) confirmed the morphology of
the PWN system with the 65 millisecond energetic young neutron star at its
core. Again the inferred spin down age, based on the existing empirical
relation between pulsar spin down power and X-ray luminosity led to a
characteristic spin down age that was considerably more than 2000 years.
Indeed on the basis of radio measurements of $\dot{P}$ for PSR J1811$-$1925, at 
$\sim$4.4$\times$ 10$^{-14}$s~s$^{-1}$ the characteristic spin down age would be
placed at more than 20,000 years.  The characteristic spin down age has also been found to be inconsistent in the following cases: Pulsar B1951+32 \citep{2002ApJ...567L.141M}; Pulsar J0538+2817 \citep{2003ApJ...593L..31K}.

\section{Summary and Conclusions}

All the evidence compiled from observations of SNR G12.8$-$0.0./IGR J18135$-$
1751/HESS J1813$-$178 points to the presence of a young and energetic
pulsar, which is powering the system. However, if we assume that this as yet
undetected pulsar was born within SNR G12.8$-$0.0 with a period that was
extremely short compared to the current and unknown value, then it is not
possible to reconcile the likely properties of the putative pulsar with the
estimated age of the SNR without significantly surpassing the physical
limits of currently known pulsar/PWN systems or stretching both the age and
the distance of the object to or beyond the limits of the accepted values
derived from current observational evidence. Such extreme configurations are
of course possible, but must be considered unlikely. If, however, we assume
that the putative pulsar was born in SNR G12.8$-$0.0 with a period that was
close to its current and as yet unmeasured value then the dilemma goes away.
The discovery of such a pulsar would be likely to reveal a period of some
tens of milliseconds. This is obviously an intriguing and unusual system that will only be understood by future observations and a detection of the putative pulsar.

\section*{Acknowledgments}

Based on observations with INTEGRAL, an ESA project with instruments and
science data centre funded by ESA member states (especially the PI
countries: Denmark, France, Germany, Italy, Switzerland, Spain), Czech
Republic and Poland, and with the participation of Russia and the USA.

We thank Dr A.~J. Bird and the INTEGRAL/IBIS survey team for access to
and the use of the data products produced in the production of the 3$^{rd}$
IBIS/ISGRI survey.

We acknowledge the funding via PPARC grant PP/C000714/1.

\label{lastpage}

\end{document}